\documentstyle[12pt,epsfig]{article}
\def\lsim{\raise0.3ex\hbox{$<$\kern-0.75em\raise-1.1ex\hbox{$\sim$}}}
\def\gsim{\raise0.3ex\hbox{$>$\kern-0.75em\raise-1.1ex\hbox{$\sim$}}}
\begin {document}
\baselineskip 24pt
\parindent=15pt
\rightline{{\bf US-FT/22-97}}
\rightline{July 1997}
\vspace{0.3cm}
\begin{center}
{\bf   PROBABILISTIC VERSUS FIELD-THEORETICAL DESCRIPTION OF HEAVY FLAVOUR
PRODUCTION OFF NUCLEI$^{a}$
}\\ \vskip .5 truecm
{\bf M. A. Braun$^{b}$,
C. Pajares and C. A. Salgado}\\
{\it Departamento de F\'{\i}sica de Part\'{\i}culas, Universidade de Santiago
de
Compostela, \\
15706-Santiago de Compostela, Spain}\\
{\bf N. Armesto$^{c}$
and A. Capella} \\
{\it Laboratoire de Physique Th\'eorique et Hautes Energies$^{d}$,
Universit\'e de Paris XI,\\ B\^atiment 211,
F-91405 Orsay Cedex, France}
\\
\end{center}

\begin{abstract}
The absorptive corrections resulting from the rescattering of a heavy flavour
 (the
so-called nuclear absorption) are usually calculated with a probabilistic
 formula valid
only in the low energy limit. We extend this formula to all energies
using a quantum field theoretical approach. For
 charmonium
and bottonium we find that the absorptive corrections in the rigorous treatment
 are
very similar to the ones in the probabilistic approach. On the contrary, at
sufficiently high energy, open charm and bottom are absorbed as much as
 charmonium and
bottonium - in spite of the fact that their absorptive cross-sections are zero,
 and therefore they
are not absorbed in the probabilistic model. At high enough energies there are
also
absorptive corrections due to the shadowing of the nucleus structure function,
 which
are present for all systems including Drell-Yan pair production. These
shadowing corrections cancel in the low energy limit.

\end{abstract}
\vfill

{\small
\noindent $^a$ Work supported by INTAS 93-79.\\
\noindent $^{b}$
Permanent address: Department
of High-Energy Physics,
University of St. Petersburg, St. Petersburg, 198904 Russia.\\
\noindent $^{c}$
Now at II. Institut f\"ur Theoretische Physik, Universit\"at
Hamburg, Luruper Chaussee 149, D-22761, Germany. \\
\noindent $^{d}$
Laboratoire associ\'e au Centre National de la Recherche
Scientifique - URA D0063.
}

\def\beq{\begin{equation}}
\def\eeq{\end{equation}}
\def\to{\rightarrow}

\newpage
\section{Introduction}
The study of rescattering effects in heavy flavour production in hadron-nucleus
 and
nucleus-nucleus collisions is of particular interest for the interpretation of
$J/\psi$ suppression found experimentally at CERN \cite{1.}.
It is well known that an
important
part of this suppression is due to the rescattering of the pre-resonant
$c\bar{c}$ system with the nucleons of the colliding nuclei. This phenomenon is
known as nuclear absorption. To describe it the following probabilistic formula
\cite{2.,3.}
is currently applied
\[
\sigma_{pA}^{\psi} = \sigma_{pN}^{\psi} \ A \int d^2b
 \int_{- \infty}^{+ \infty} d z
\  \rho_A (b, z) \exp \left ( -\sigma_{abs}^{\psi} \ A
\int_{z}^{+ \infty} d z' \ \rho_A (b, z') \right )\]\beq=
\sigma_{pN}^{\psi} (\sigma_{abs}^{\psi})^{-1} \int d^2b \left
[ 1 - \exp \left ( - \sigma_{abs}^{\psi} A T_A(b) \right ) \right ] \ \ \ .
\eeq
In this formula $\rho_A$ and $T_A$ are the total and transverse nuclear
densities respectively, $\sigma_{pA(N)}^{\psi}$ are the inclusive $J/\psi$
cross-section on $A(N)$ and $\sigma_{abs}^{\psi}$ is the absorptive
 $(c\bar{c})$-$N$
cross-section (i.e. corresponding to final states without $J/\psi$). For open
 heavy
flavour production $\sigma_{abs} = 0$ and one obtains an $A^1$ behaviour in
 agreement
with experiment \cite{4.}.
For charmonium and bottonium $\sigma_{abs} \not= 0$ and one
obtains a behaviour $A^{\alpha}$ with $\alpha < 1$ - also in agreement with
experiment \cite{1.,5.}.
Eq. (1) has a probabilistic interpretation with a clear
longitudinal ordering in $z$: in the first interaction at $z$
the heavy system is produced and in successive ones at $z' > z$
it
rescatters with nucleons along its path. However, there is a fundamental change
in the physical picture as one goes from low energies, for which
the probabilistic formula (1) is derived, to high energies.
Instead of successive interactions of the projectile with
nucleons of the target, one has simultaneous
interactions of particles into which the
projectile has split. Thus the nuclear interaction responsible
for the heavy particle production may occur after the interactions
responsible for its absorption in nuclear
matter. To illustrate this somewhat paradoxical point, compare Figs. 1a and 1b,
 in
which the heavy particle production and its subsequent
interactions is presented as seen in the lab. system. The time goes from left
to
 right.
It is clear that in both cases the heavy particle is in fact created long
before
 the
actual interactions with the nucleus. In Fig. 1a the times of these
interactions
correspond to the common idea of production first, absorption after. However in
 Fig. 1b
the order of interactions is the inverse one: the interaction of the
(already formed) heavy
particle precedes the one which is responsible for its formation. \par

In what follows we are going to study the changes in Eq. (1) resulting from
this
change in the physical picture at high energy. We shall derive an expression
 valid at
all energies which exactly coincides with (1) in the low energy limit. \par

Our expression contains not only the rescattering of the heavy
 system
 but also the rescattering of light particles (gluons and
light quarks).  The latter can be interpreted as shadowing
corrections to the nucleus structure function and cancel
out in the low energy limit. Also our formula splits into
pieces with no interactions of  light  particles, which
we call the internal components and which are shown to be
small in the central rapidity region,
and a piece with at least one light particle
interaction, which we call external and which corresponds to
the usual production mechanisms such as gluon-gluon fusion.

At asymptotic energies the rescattering of the heavy flavour in the external
 component
has a form different from (1). The change consists in the substitution
\beq(\sigma_{abs}^{\psi})^{-1} \left [ 1 - \exp \left ( -\sigma_{abs}^{\psi} \
A
 \ T_A(b)
\right ) \right ] \longrightarrow
A\ T_A(b) \exp \left ( - {1 \over 2} \widetilde{\sigma} \
 A \
T_A(b) \right ), \eeq
\noindent where $\widetilde{\sigma}$ is the $c\bar{c}-N$ total cross-section.
It
 is
most interesting that for $\widetilde{\sigma} \simeq \sigma_{abs}$, which is
the
 case
for charmonium or bottonium production, the first rescattering correction (i.e.
 the
term proportional to $\widetilde{\sigma}$) is the same in the two cases.
 Since
$\widetilde{\sigma}$ is small the change with increasing energy in the
 absorptive
corrections due to the rescattering of the heavy system will be comparatively
 small. As a consequence, the probabilistic expression remains approximately
valid even at LHC energies
(if the shadowing of the nuclear structure functions is
neglected).

The situation is completely different for open charm or bottom production. In
 this
case $\sigma_{abs} = 0$ and, as we said before, (1) leads to $A^1$. However
with
increasing energy there are shadowing corrections resulting from the
 rescattering of
the heavy system (despite $\sigma_{abs} = 0$). At high energies ($\sqrt{s}\geq$
200GeV) they are
practically identical to those for the $J/\psi$. This is a main prediction of
 our
approach.

On top of these effects due to the rescattering of the heavy system,  there
are,
of course, corrections due to the
 shadowing in the nucleus structure functions. The latter are present at
high enough
 energy for
all systems including Drell-Yan pair production, but vanish in the low energy
limit.

The paper is organized as follows. In Section 2 we derive the expressions valid
 at
asymptotic energies which show the change (2) in the absorptive series for the
external component. These formulae are in fact applicable only at extremely
high
energies. Indeed, due to the presence of the heavy system there are finite
 energy
corrections which are important up to energies of the order $M^2R_A/x_+$, where
 $M$  is
the mass of the heavy system, $x_+$ its light cone momentum fraction and $R_A$
 the
nuclear radius \cite{6.,7.}.
These finite energy corrections are explicitly computed
 in
Section 3. In this Section we also show that, in the low energy limit, our
 formulae
coincide exactly with the probabilistic formula (1). Section 4 contains our
numerical results. Our conclusions are summarized in Section 5.

\section{Heavy particle production at asymptotic energies}
In this Section we study the inclusive cross-sections for the
production of a heavy flavour system in a high-energy
 collision, including the rescattering of the heavy
system. We start with  $pA$ collisions. The total cross-section
 for production of the heavy
system can then be represented by the diagram shown in Fig. 2
\cite{8.}. It has a clear meaning in the laboratory reference system: at
some point the heavy system is born from light partons
(gluons), whereafter both heavy and light particles scatter on
the nucleus. In the probabilistic treatment leading to Eq.
(1) there is a time ordering: first an interaction of a
light particle occurs, in which the heavy system is
created, and only afterwards the latter interacts with the
nucleus.  As explained in the Introduction this time ordering
does not exist at high energies where all interactions of
light and heavy particles are simultaneous. We shall treat
interactions with the nucleus of both the light particle  and
heavy system in the eikonal approximation. We also take all
participating particles as scalar, for simplicity.

\subsection{The external contribution}

Using an eikonal model for the multiple scattering (both of light and heavy
systems) in Fig. 2b, we obtain for the external contribution to the inclusive
cross-section of a heavy system, integrated over the
transverse momenta:
\beq
I^{(ext)}_{A}(x)=\frac{\pi g^{2}}{
M^{2}}F_{p}(x_{1},M^{2}) F_{N}(x_{2},M^{2})\Phi_{A},
\eeq
where
\beq
\Phi_{A}=\frac{1}{\sigma}[ \sigma_{ A}(a+\tilde{a})
 -\sigma_{ A}(\tilde{a})] \ \ \ . \eeq
Here $gM$ corresponds to the hard scattering vertex, $M$ is the heavy particle
mass, $x_{1}$ and $x_{2}$ are the longitudinal momentum fractions of the
projectile and target carried by the colliding partons,
$x=x_{1}-x_{2}$,
 $M^{2}=x_{1}x_{2}s$ where $s$ is the overall
energetic variable: $s=(p_{1}+p_{2})^{2}$, and
$ p_{2} $ is the momentum of a nucleon in the target nucleus.
$F_{p(N)}(x_{1},M^{2})$ is the structure function of the
projectile proton (nucleon) at $Q^2 = M^2$. $a$ is the light particle-nucleon
amplitude, with $\sigma = 2 \ {\rm Im} \  a$. The
 notation $\sigma_{ A}(a)$ means the total
cross-section calculated with the
amplitude $a$: \beq
\sigma_{A}(a)=2\ {\rm Re}\int\,d^{2}b\left(
1-[1+iaT_{A}(b)]^{A}\right) \ \ \ ,
\eeq

\noindent
with $T_A(b)$ the nuclear preofile function normalized to 1.
Finally, we have introduced the amplitude
$ \tilde{a} $ for the scattering of the on-mass-shell heavy
system
on a nucleon from the target nucleus\footnote{In the
case when the heavy particle is either a charmonium or
bottonium bound state there are two different time scales
involved, the production of the heavy system containing the
$c\bar{c}$ or $b\bar{b}$ pair and that of the bound state.
Although we make no distinction in what follows between heavy
system and heavy particle it has to be understood that what
propagates through the nucleus and interacts with the nucleons
is not the bound state but the pre-resonance heavy system which,
for convenience, we denote $ \Psi $ in the following.}, with the
cross-section $\widetilde{\sigma} = 2 \ {\rm Im} \  \widetilde{a}$.

An explicit derivation of Eqs. (3) and (4) is given in Sect. 3. Expressions
(3)-(4) have a simple physical meaning. Note first that for the production on a
single nucleon they reduce to

\beq
I^{(ext)}_N(x) = {\pi g^2 \over  M^2} F_p(x_1, M^2) \ F_N(x_2, M^2) \ \ \ .
\eeq

\noindent  This expression corresponds to the standard QCD approach in which
the
heavy
flavoured system is produced in a hard collision of two gluons. The nuclear
effects are just given by the total cross-section $\sigma_A(a + \widetilde{a})$
with an amplitude equal to the sum $a + \widetilde{a}$. It corresponds to the
scattering
on the nucleus of a ``projectile'' containing light and heavy particles.
The negative term in Eq.
(4) eliminates the contribution with no interaction of the
light particle, which, according to the definition given in the Introduction,
is
part of the internal
contribution. As we shall see below the internal contributions
have to be treated differently and turn out to be numerically small at
mid-rapidities.

In order
to see more clearly the physical meaning of Eqs. (3) and (4), we write them at
fixed impact parameter, assuming that the amplitudes $a$ and $\widetilde{a}$
are
purely imaginary. Eqs. (3) and (4) reduce in this case to

\beq I_A^{ext}(x , b) = {2\pi g^2 \over M^2} F_p(x_1, M^2) F_A(x_2, M^2,
b) e^{-{1 \over 2} \widetilde{\sigma} AT_A(b)}, \eeq

\noindent where

\beq F_A(x_2, M^2, b)
= F_N(x_2, M^2) {1 \over \sigma} \left [ 1 - e^{-{1 \over 2}
\sigma A T_A(b)} \right ] \eeq

\noindent has
the meaning of the structure function of the target nucleus at impact
parameter $b$ (see below). The last factor in Eq. (7) corresponds to the
rescattering
of the heavy system. By comparing with the corresponding rescattering
in the probabilistic
approach, Eq. (1), we obtain the substitution (2) mentioned in
the Introduction if we neglect the shadowing corrections to the nuclear
structure functions coming from (8).
For $J/\psi$ production, $\widetilde{\sigma} \simeq \sigma_{abs}$
and the first absorptive correction (i.e. the term proportional to
$\widetilde{\sigma}$) is the same in the probabilistic approach and in the
asymptotic formula. Since $\sigma_{abs}$ is small, the absorptive corrections
to
$J/\psi$ production
due to the rescattering of the heavy system will be similar in
the two cases
(see Sect. 4 for a more detailed discussion).  This result, which was
first anticipated in Ref.
\cite{6.}, is by no means trivial.  Indeed if one would
naively remove the time ordering in the probabilistic expression (performing
the
integral in $z$ from $- \infty$ to $+ \infty$) one would obtain an absorptive
correction of the form (7) but with an exponent two times larger.

Another important point is that (7) contains the total cross-section
$\widetilde{\sigma}$ instead of the absorptive part $\sigma_{abs}$.
As a result, the situation is very different
for charmonium and for open charm production. In the latter case $\sigma_{abs}
=
 0$
and thus the probabilistic expression gives an $A^1$ behaviour. This result
will
 also
be obtained from our field theoretical treatment in the low energy limit.
 However, at
high energies, as seen from (7), there will
be absorptive corrections of  the same magnitude as for
charmonium. This is an interesting prediction of our approach.

In concluding this Subsection we would like to emphasize that the shadowing
corrections to the nucleus structure function Eq. (8), have been described by
an
eikonal formula or, more precisely, by a factorized multi-Pomeron vertex. We
could have used instead the sum of tree diagrams with the triple Pomeron
coupling
(eikonalized Schwimmer formula) or a pure eikonal formula.
The smallness of the shadowing corrections in these approaches is due to
the weakness of the triple Pomeron coupling or to off-mass-shell effects
(in the latter case). Our treatment can certainly be generalized to
include the pure eikonal mechanism and, possibly, other mechanisms
as well. However
 our aim in this work is not to describe in detail these
shadowing corrections but rather
the nuclear effects due to the rescattering of the
heavy system. So we do not elaborate on possible generalizations
as far as the EMC effect at low $x$ is concerned.
This effect is small at low energies and vanishes in the low-energy limit.
 This result,
which will be derived explicitly in Sect. 3, is due to the fact that at low
energies the momentum fraction $x_2$ associated to the target is large -
outside
the shadowing region. Technically, in the formalism of Sect. 3, this
vanishing is due
to $t_{min}$ effects.

\subsection{The internal component}

We turn next to the internal component. As defined in the Introduction the
internal component of the projectile (target) corresponds to the case when the
heavy system is produced from partons belonging to the projectile (target)
and no
interaction takes place between light partons of projectile and target.
It turns out
that in the central region this component is very small at high energies.
At low energies it is also
small for charmonium. For open charm it is not small but is absorbed exactly in
the same way as the external component. Therefore,
in the central region one would obtain the same
results for the $A$-dependence by neglecting this internal component.
In the projectile fragmentation region,  the internal component from the
target becomes dominant, since other components evidently vanish in the limit
$x_F\rightarrow 1$. To calculate it correctly, however, one has to properly
take into account shadowings corrections to the nuclear structure function
at extremely low $x$, which are presumably very important.
This will be done in our numerical calculations as discussed in Sect. 4.

 The internal contribution evidently involves
the probability to find the heavy system with a given scaling
variable
$ x'_{1} $ in the projectile - described by the part of the projectile
structure function contributed by the heavy flavoured particle
$F_{\Psi/p}(x'_{1},Q^{2})$.  Since
the  heavy
system formed in the projectile may have an arbitrary scaling
variable $ x'_{1}\geq x_{1} $,  integration over
$ x'_{1} $ is expected. Upon colliding with the nucleus the
heavy system with the scaling variable $ x'_{1} $ has to
transform into the observed heavy system with the scaling
variable $ x_{1} $. This process is described by the inclusive
cross-section $I_{\Psi A\rightarrow\Psi}(x'_{1}\rightarrow x_{1})$.
 From the diagrams in Fig. 2 one finds
 \beq
I^{(int,p)}_{A}(x_1)=\int
_{x_{1}}^{1}(dx'_{1}/x'_{1})F_{\Psi/p}(x'_{1},M^{2}) I_{\Psi
A\rightarrow\Psi}(x'_{1}\rightarrow x_{1}) \ \ \ . \eeq

The heavy particle component of the structure function can be
calculated via the light particle one assuming that the heavy particle
is produced inside the projectile by a hard scattering mechanism  (i.e.
that there is no intrinsic heavy particle component).
In our simplified scalar case we find
 \beq
F_{\Psi/p}(x,Q^{2})=\frac{g^{2}x^2}{16\pi^{2}}
\int_{x}^{1}\frac{dx_{1}}{x_{1}^{3}}
F_{p}(x_{1},Q^{2}) \ \ \ .
\eeq
 From (10) one concludes that the heavy flavour structure
function is smaller than the ordinary one by a factor $g^{2}$ (actually taken
at the scale $Q^{2}\sim M^{2}$ and therefore small). It is also damped in the
region $x\sim1$ by an extra power of $(1-x)$.

As to the inclusive cross-section
 $I_{\Psi A\rightarrow\Psi}(x_{1}\rightarrow x) $,
we can estimate it knowing that in the rescattering the heavy
system tends to conserve its longitudinal momentum. Then we approximately have
 \beq
I_{\Psi A\rightarrow\Psi}(x_{1}\rightarrow x)=x_{1}
\sigma_{\Psi A}\delta(x-x_{1}) \ \ \ .
\eeq
This simple result is however valid only for
open heavy flavour ($D\bar{D}$, etc.). For hidden heavy flavour (e.g.
$ J/\psi $) one expects an extra absorption in each
inelastic collision due to transition into  open heavy
flavour channels. To describe this absorption we introduce a
factor
$ \varepsilon $ which represents the part of the initial
heavy flavoured cross-section which contains the observed
heavy system. For open flavour
$ \varepsilon=1 $ and for charmonium or bottonium
$ \varepsilon \ll 1 $. After introducing
$ \varepsilon $, (11) changes to
 \beq
I_{\Psi A\rightarrow\Psi}(x_{1}\rightarrow x)=x_{1}\delta(x-x_{1})
\sum_{n=0}^{A}
\epsilon^{n}\sigma_{\Psi A}^{(n)} ,
\eeq
where $\sigma_{\Psi A}^{(n)}$ denotes the
$ \Psi$-$ A $ cross-section with
$ n $ inelastic interactions.

Putting (12) into (9) we obtain the internal contribution from the
projectile in
a simple form (an explicit derivation is given in Sect. 3): \beq
I^{(int,p)}_{A}(x)= \sigma^{(\varepsilon)}_{\Psi A}F_{\Psi/p}(x_{1},
M^{2}) ,
\eeq
where
\beq
\sigma^{(\varepsilon)}_{\Psi A}=
\sigma^{(el)}_{\Psi A}(\tilde{a})+ \sigma^{(in)}_{ \Psi A}(\tilde{a})-
\sigma^{(in)}_{\Psi  A}(\tilde{a}\rightarrow\tilde{a}(1-\varepsilon))
\eeq
and  $\sigma^{(in)}_{ \Psi A}(\tilde{a})$ is the $ \Psi A $
inelastic
 cross-section  calculated with the
$\Psi$-$ N$ amplitude $\tilde{a}$. For open flavour production
($\varepsilon=1$) we evidently have
$\sigma^{(\varepsilon)}_{\Psi A}=\sigma^{(tot)}_{\Psi A}$, whereas for hidden
flavour with large absorption ($ \varepsilon\simeq 0 $)
$\sigma^{(\varepsilon)}_{\Psi A}
\simeq \sigma^{(el)}_{\Psi A}$. In the last case the internal contribution from
the
projectile is therefore substantially reduced.

Apart from  the internal flavour in the projectile,
we have also to take into account the internal flavour
in the target nucleus, which upon scattering off the projectile will
also contribute to the cross-section. Its expression can be
easily obtained. We get
\beq I^{(int,A)}_{A}(x)=\varepsilon\tilde{\sigma}
F_{\Psi/N}(x_{2}, M^{2})\Phi_{A} ,
 \eeq
where $ \tilde\sigma $ is the $
\Psi N $ cross-section and the factor $\Phi_{A}$ is given by (4).

For the production on a single nucleon we find
 \beq I^{(int,p)}_{N}(x)=\varepsilon\tilde{\sigma}
F_{\Psi/p}(x_{1},
M^{2}) \eeq
and
 \beq I^{(int,N)}_{N}(x)=\varepsilon\tilde{\sigma}
F_{\Psi/N}(x_{2},
M^{2}) \ \ \ . \eeq

To have a more explicit form
we rewrite (16) and (17) for fixed impact parameter
$ b $.

\noindent
Assuming that $a$ and $\widetilde{a}$ are purely imaginary and restricting
ourselves to the case of open flavour
 ($ \varepsilon=1 $) we have \beq
I^{(int,p)}_{A}(x,b)=2
F_{\Psi/p}(x_{1},M^{2})
  (1 - e^{- (1/2){\tilde\sigma}A  T_A(b)})
\eeq
and
\beq
I^{(int,A)}_{A}(x,b)=\tilde{\sigma}
F_{\Psi/A}(x_{1},M^{2},b)
  e^{- (1/2){\tilde\sigma}A  T_A(b)} ,
\eeq
where $F_{\Psi/p}(x_{1},M^{2},b)$ is the heavy flavoured part of the
nucleon structure function, Eq. (10), at fixed $b$, and
\beq
F_{\Psi/A}(x_{2},M^{2},b)= F_{\Psi/N}(x_{2},M^{2})\frac {1}{\sigma}
  [1 - e^{- (1/2)\sigma A  T_A(b)}] \ \ \ .
\eeq

As it was the case for the external component, (18) and (19) actually describe
two
different physical effects: a change of the gluon distribution in the
colliding nucleus as compared to the free nucleon (shadowing part of the EMC
effect) and the rescattering of the heavy system in the nucleus
target. The former effect corresponds to terms  of the
 second and higher powers in the light-particle interaction
 with the target, i.e.  in the amplitude $ a $  or
 cross-section $ \sigma $. Rescattering of the heavy system is described by
 the last factor in (18) and (19).

Note that the internal part coming from the projectile, Eq. (18),  is absorbed
in a
different manner. It is proportional to
the total cross-section for $\Psi$-$A$
scattering and behaves like $A^{2/3}$ whereas
the two other parts behave like $A^{1/3}$ (at very large $A$). However, as we
shall see numerically, both internal contributions are much smaller than the
external one at $x_F\sim 0$.

In concluding this Subsection we would like to emphasize that, while the
internal
contribution does contain $\varepsilon$, the external one does not. This is due
to
the following.
In order to compute the inclusive cros-section for the production of
the heavy system we have to fix an intermediate
on-mass-shell heavy system in Fig. 2 (i.e. ``cut" the diagram
through the heavy system line).
This cut may pass either through the rescattering blob $A$ or through one of
the two heavy particle lines with which the blob is attached to the rest of the
diagram. If the cut passes through the blob $A$ then the lower blob $B_{t}$ may
be both cut or uncut. The corresponding contributions  cancel \cite{8.} in
this sum due to the well-known AGK
cancellations \cite{9.}. Therefore we have
to consider only the two cases when the cut passes either
through the left or through the right heavy particle line in
Fig.  2b. Due to this important result
the blob $ A $, containing the
inclusive cross-section of the heavy system, is never cut -
and thus $\varepsilon$
never appears.
However, this AGK cancellation is only true for the external component.
Clearly,
it does not take place in the case when the light particle
does not interact with the nucleus at all. Therefore the internal contribution
does
involve the cut blob $ A $ and has to be treated separately. In particular it
contains the factor $\varepsilon$.

\subsection{Generalization to nucleus-nucleus collisions}

In spite of a considerable
complication in the dynamics, it can be shown that the AGK cancellation which
governs the contributions to the inclusive cross-sections
discussed in the last paragraph of Sect. 2.2 remains valid for $AB$ collisions.
Namely, emission of the heavy system from  the rescattering blob is
cancelled out, unless there is no interaction of the light
particle with one of the nuclei. As a result, in the $AB$ case
we have to consider diagrams of the same type as for $hA$
scattering and, correspondingly, the contributions to the heavy system
inclusive
cross-section are divided into an external part, with
interactions of at least one light particle of each nucleus with nucleons of
the
other nucleus, and two internal parts: that of the nucleus $B$, with no
light particle interaction with nucleus $A$, and that of the nucleus $A$,
with no
light particle interaction with nucleus $B$. All three parts
are calculated quite similarly to the $hA$ case. We present here
only the final results for the inclusive cross-sections.

For the external part we find an expression which is a
generalization of Eqs. (3) and (4):
\beq
I^{(ext)}_{A}(x)=\frac{\pi g^{2}}{
M^{2}}F_{N}(x_{1},M^{2}) F_{N}(x_{2},M^{2})\Phi_{A}\Phi_{B} \ \ \ .
 \eeq
  As one observes, the inclusive cross-section (21)
is factorized in the two nuclei, $A$ and $B$. Recalling (3)  and
(6) it can be rewritten as a simple relation \beq
\frac{I_{AB}^{(ext)}}{I_{N}^{(ext)}}=
\frac{I_{A}^{(ext)}}{I_{N}^{(ext)}}\ \frac{I_{B}^{(ext)}}{I_{N}^{(ext)}} ,
\eeq
where in the $hA$ and $hB$ collisions the nuclei are
taken in the same kinematics as in the $AB$ collision.
Evidently (22) means that the total absorption is just
the product of absorptive factors coming from both nuclei.
As is well-known this relation also holds in the probabilistic
approach \cite{2.,3.}.

For the internal parts we also obtain expressions factorized
in the two colliding nuclei. The internal heavy
flavour present in the nucleus $B$ gives a contribution
which generalizes Eqs.(13) and (14):
\beq
I_{AB}^{(int,B)}(x)=\sigma_{\Psi A}^{(\varepsilon)}
F_{\Psi/N}(x_{1},M^{2})\Phi_{B} \ \ \ .
 \eeq
To this internal contribution a similar one
$I_{AB}^{(int,A)}$ has to be added, which takes into account the
internal heavy flavour of the target. It is given by (23) with the exchanges $A
\leftrightarrow B$ and $x_1 \leftrightarrow x_2$.

\def\tia{i\tilde{a}}
\def\tis{\tilde{\sigma}}

\section{
 Finite energy corrections}

\subsection{The formalism}

As discussed in the Introduction, finite energy effects are very important in
 heavy
flavour production. These effects are present up to rather high energies
 especially at
small $x_F$ (see below). In this Section we are going to consider these
 corrections.
By introducing them we obtain equations valid at all energies.
At asymptotically high energies they coincide with the ones
derived in the previous Section. In the low energy limit,
they coincide with the result of the probabilistic approach,
Eq.  (1).  As discussed in Refs. \cite{6.} and \cite{7.} the finite
energy corrections have a clear origin. Due to the presence of
the heavy system, some of the contributions to the inclusive
cross-section have a non-vanishing minimal transverse momentum
($t_{min} \not= 0$) and are suppressed by the nuclear form
factor in a well-defined way. These  modified cutting rules
have been computed in Ref. \cite{7.} in the framework of a specific
parton model. However, their physical content is so
transparent that they presumably have a more general validity.
They can be summarized in the following way. Let us
consider a particular ordering of the longitudinal coordinates
$z_j$ of
$ n$ interactions with the nucleus of the light particle and
the heavy system, $n = l+ h$, where
$ l(h)$ is the number of the interactions of the light
particle (heavy system) and  \beq z_1 \leq z_2 \leq \cdots
\leq z_n \ \ \ .  \eeq
 In this case, at finite energies,
the
$n$-th power of the nucleus profile function $T_A^n$ which
appears in the expansion in the number of interactions of the cross-section
$ \sigma_{A} $, Eq. (4),
has to be replaced \cite{7.}
by one of the following integrals: \beq
T_{n}^{(j)}(b)=n!\int_{-\infty}^{+\infty}dz_{1}\int_{z_{1}}^{+\infty}dz_{2}...
\int_{z_{n-1}}^{+\infty}dz_{n}\exp(i\Delta
(z_{1}-z_{j}))\prod_{i=1}^{n} \rho_{A}(b,z_{i}), \eeq
where
$ j=1,2,...,n $
 and
 \beq\Delta =  m_N M^2/s \ x_+ \ \ \ .
\eeq  Note
that for $\Delta = 0$, corresponding to asymptotic
energies, all integrals $T_n^{(j)}$ are equal to $T_A^n$ and
are independent of $j$. \par

For $\Delta$ non zero the value of $j$ to be taken in Eq. (25)
depends on the particular discontinuity of the scattering amplitude
 we are considering. Namely, all the discontinuities
containing $T_n^{(j)}$ are of the following type.
Interactions with the nucleus from 1 to
$ j-1 $ have to be located to the left of the considered cutting line (``cut to
the
right"). Interaction
$ j $ may either be cut or be located to the right of the cutting line
(``cut to the left"). All the other interactions, from
$ j+1 $ to
$ n $ may be cut in all possible ways. To this contribution
 one has to add its complex conjugate.

The physical content of these rules is quite clear. When the first interaction,
in the ordering (24), is cut the exponential damping factor in
(25) is not present, i.e. $T_n^{(1)} = T_A^n$. Clearly this
is the only case where $t_{min} = 0$ \cite{6.}. In all other cases
the exponential damping factor is present and depends on the
longitudinal distance $z_j - z_1$ between the first
interaction 1 and the first one $j$ which is either cut or
cut to the left \cite{7.}.

Turning to the inclusive cross-sections, we have to stress that
for
$ \Delta>0 $ the AGK cancellation discussed at the end of Sect. 2.2 is no more
valid. Therefore we have to consider all possible cuttings on the same footing.

In treating the emission from the rescattering blob we apply the
same approximation as was used in Sect. 2 for the internal
contribution from the projectile. Namely we assume that the
 observed heavy system
conserves the original longitudinal momentum. As in Sect. 2 we
introduce the factor $ \varepsilon\leq 1 $ for each inelastic
interaction of the heavy system to take into account a
possible leakeage of the hidden flavour into the open one.

Using these simple rules it is now easy to write the
expression for the inclusive cross-section at finite energies
(i.e.  $\Delta \not= 0$). In order to do so one has to
remember that when cutting an interaction to the right
(left) the amplitude $ia$ is replaced by $ia$ ($- ia^*$),
irrespective of whether the particle is light or heavy.
Cutting an interaction of a light particle one has to replace
$ia$ by $\sigma$.  Cutting that of a heavy particle one has to
replace $i\widetilde{a}$ by $\varepsilon \widetilde{\sigma}$.

The expression for the external part of the inclusive
cross-section can  then be written in the same form (3),(4), where
now
\beq
\sigma_{A}(a+\tilde{a})-\sigma_{A}(\tilde{a}) \longrightarrow
 \widehat{\sigma}_A (a, \widetilde{a}) - \widehat{\sigma}_A
(a = 0, \widetilde{a}),
 \eeq
 with

\beq
\widehat{\sigma}_A (a, \widetilde{a}) = {\rm Re} \ \sum_{n=1}^A C_A^n
\sum_{j=1}^n
\int d^2 bT_n^{(j)} (b) \sigma_n^{(j)}(b) \ \ \ .
\eeq

Here for $j > 1$

\beq\sigma_n^{(j)}(b) =  2 \
(ia + i \widetilde{a})^{j-1} (-ia - i \widetilde{a} - (1 -
\varepsilon ) \widetilde{\sigma}) [- (1 - \varepsilon )
\widetilde{\sigma}]^{n-j} \eeq
 and for $j = 1$
\beq\sigma_n^{(1)}(b) =  (\sigma +
\varepsilon \widetilde{\sigma}) [- (1 - \varepsilon)
\widetilde{\sigma} ]^{n-1} \ \ \ .  \eeq
 Eq.
(29) is quite obvious: the factor $(ia + i
\widetilde{a})^{j-1}$ results from the cutting to the right of
the first $j - 1$ interactions; the second factor corresponds
to the cutting of the
$ j $-th interaction and the last one to
the cutting of interactions from $j + 1$ to $n$.
The first factor is trivial since a right cutting does not
change the amplitudes $ia$ and $i\widetilde{a}$.
The  cutting of the interaction plus its cutting to the left
 that appears in the second factor leads to $-ia$
for a light particle and to $-i \widetilde{a} - (1 -
\varepsilon ) \widetilde{\sigma}$ for a heavy one.
Indeed, the cutting
of a light particle interaction in all possible ways (right +
left + interaction itself) gives $ia - ia^* + \sigma = 0$.
The cutting of a heavy system interaction in all possible ways
gives $(\varepsilon - 1)\widetilde{\sigma}$. This also explains the
last factor of (29).

Eq. (30) is obtained after putting $j =
1$. In this case the $j$-th interaction is the first one, and thus it has to be
cut either to the left or through its interaction. It can be seen that the
former
(cutting to the left) is already included in the complete conjugate
 term responsible for the
factor 2 in (29). Only the
cut interaction remains in this factor. No complex conjugate term appears in
 this case.

For the internal part from the projectile,
 applying the same rules, we obtain Eq.
(13) where the cross-section
$ \sigma_{\Psi A}^{(\varepsilon)} $ is now replaced by $\widehat{\sigma}_{\psi
A}^{(\varepsilon )}$, with  \beq
 \widehat{\sigma}_{\Psi A}^{(\varepsilon)}=
{\rm Re}\,\sum_{n=1}^{A}C_{A}^{n}\sum_{j=1}^{n}
\int d^{2}bT_{n}^{(j)}(b)\sigma_{n}^{(j)}\ \ ,
\eeq
where for $j>1$
\beq
\sigma_{n}^{(j)}=2 (\tia)^{j-1}(-\tia-(1-\varepsilon)\tis)
[-(1-\varepsilon)\tis)]^{n-j}
\eeq
and for $j=1$
\beq
\sigma_{n}^{(1)}=\varepsilon\tis
[-(1-\varepsilon)\tis)]^{n-1} \ \ \ . \eeq

The internal part coming from the target contains the same rescattering
diagrams as the external part. This resulted in the same rescattering
 factor at asymptotic energies, (cfr. Eqs. (3) and (15)). Therefore at finite
energies it is given by the  same Eq. (15) with the substitution (27) in
 $\Phi_A$.

Since, by construction, each possible discontinuity has been included in one
and
 only
one term of (28)-(33), it is clear that the above formulae provide an explicit
derivation of the asymptotic energy results given in Sect. 2. Let us consider
the asymptotic case when  $\Delta = 0$ and $T_n^{(j)} = T_A^n$. It can
be easily checked that one recovers the results of that Section.
Take Eq. (28). One
can notice that, with
$ T_{n}^{(j)} = T_A^n$,
there is a cancellation between the term of $\sigma^{(j)}$
proportional to $-i a - i \widetilde{a}$ and the term of
$\sigma^{(j+1)}$ proportional to $- (1 - \varepsilon
)\widetilde{\sigma}$. As a result, in the sum over $j$, one is
left with the term proportional to $-ia - i\widetilde{a}$ from
$\sigma^{(n)}$, the one proportional to $- (1 - \varepsilon
)\widetilde{\sigma}$ from $\sigma^{(2)}$ and the term with $j
= 1$.  The first term gives $\sigma_{A}(a + \widetilde{a})$
and the sum of the other two terms gives $-\sigma_{A}^{(in)} ((1 -
\varepsilon ) \widetilde{\sigma})$. Subtracting $\widehat{\sigma}_A (a = 0,
\widetilde{a})$ (see Eq. (27)) one obtains Eqs. (3) and (4).
Likewise one can show that
for $ \Delta=0 $ one recovers Eqs. (13) and (15) for the internal components.

Finally we shall study the low energy limit by taking $\Delta \to \infty$. In
 this
case only  terms with $j = 1$  survive in Eqs. (28) and (31).
We obtain from (30), for the external contribution, \beq
I_{(A,\Delta \to
\infty)}^{(ext)}(x) =\frac{\pi g^{2}}{ M^{2}}
F_{p}(x_{1},M^{2}) F_{p}(x_{2},M^{2})\Phi_{A}^{(0)} \ \ \ .
\eeq
 For the internal contribution of the projectile we obtain from (33)
\beq
I_{(A,\Delta \to
\infty)}^{(int,p)}(x)=\varepsilon\tilde{\sigma}
F_{\Psi/p}(x_{1},M^{2})\Phi_{A}^{(0)}
\eeq
and  for the contribution of the internal
heavy flavour of the target nucleus we get from (30)
\beq
I_{(A,\Delta \to
\infty)}^{(int,A)}(x)=\varepsilon\tilde{\sigma}
F_{\Psi/p}(x_{2},M^{2})\Phi_{A}^{(0)} \ \ \ .
\eeq
In all cases
\beq
\Phi_{A}^{(0)}=
\frac{\sigma_{A}^{(in)}
((1-\varepsilon)\widetilde{\sigma})}{(1-\varepsilon)\widetilde{\sigma}}
\ \ \ . \eeq
Two important observations can be made upon inspection of Eqs.
(34)-(37). First, in the cross-sections (34) and (36) only terms
linear in $a$ (or $\sigma$) have survived. This means
that in the low energy limit the screening corrections to the
nuclear structure function disappear, as expected. Second,
comparison of Eqs. (34)-(36) shows that the absorptive
corrections for all parts of the inclusive cross-section,
external and internal, turn out to be the same in the low
energy limit.  This allows to write for the total inclusive
cross-section on the nucleus in this limit \beq
 I_{(A,\Delta \to \infty)}^{(tot)}(x)=
I_N^{(tot)}(x)\Phi_{A}^{(0)}=I^{(tot)}_N(x)
\frac{\sigma_{A}^{(in)}((1-\varepsilon)\tilde{\sigma})}{(1-\varepsilon)\tis} \
\ \ .
\eeq
Since
 $\sigma_{abs} = (1 - \varepsilon ) \widetilde{\sigma}$ we
recover \underbar{exactly} the probabilistic expression (1). As far as we know
 this
is the first time that this expression has been derived in a field theoretical
approach.

\subsection{Computational methods}

As we see from the above formulae, in order to compute nuclear effects at
finite
energies we have to deal with the new profile functions
$T_{n}^{(j)}$ which depend on the longitudinal order of the collisions and
involve the parameter $\Delta$ (Eq.  (26)).  At first sight it
seems an impossible task due to multiple longitudinal
integrations in (25). However the situation improves if we
take a Fourier transform with respect to $\Delta$. Then
instead of $T_{n}^{(j)}$ we find integrals \[
F_{n}^{(j)}(b,\xi)=\int\frac{d\Delta}{2\pi}T_{n}^{(j)}(b,\Delta)
\exp (i\Delta\xi)=\]\beq
n!\int_{-\infty}^{+\infty}dz_{1}\int_{z_{1}}^{+\infty}dz_{2}...
\int_{z_{n-1}}^{+\infty}dz_{n}\delta (z_{1}-z_{j}+\xi)\prod_{i=1}^{n}
\rho_{A}(b,z_{i}) \ \ \ .
\eeq
Evidently $F_{n}^{(j)}(\xi)=0$ for $\xi<0$ and
\beq
F_{n}^{(1)}(\xi)=\delta (\xi)T^{n}\eeq
(here and in the following we suppress the argument
$ b $ and the subindex
$ A $).
For $j>1$ and
$\xi>0$ simple calculations lead to
\beq
F_{n}^{(j)}(b)=\frac{n!}{(n-j)!(j-2)!}
\int_{-\infty}^{+\infty}dz\rho(z)\rho(z+\xi)T^{j-2}(z,z+\xi)T^{n-j}(z+\xi) ,
\eeq
where \beq
T(z)=\int_{z}^{+\infty}dz'\rho(z')
\eeq
and \beq
T(z_{1},z_{2})=\int_{z_{1}}^{z_{2}}dz'\rho(z')=T(z_{1})-T(z_{2}) \ \ \ .
\eeq
Thus the calculation of all nontrivial $F_{n}^{(j)}$ reduces to the
one-dimensional integral (41).

Moreover, using the representation (41), one can perform the sums
over $n$ and $i$ in the formulae for the inclusive cross-sections
(28) and (31) and thus obtain for them closed expressions as a function
of $\xi$. We restrict ourselves to the pure rescattering contributions, that is
to
terms linear in
$ a $.

For the external contribution  we then  obtain
\beq
I_{A}^{(ext)}=I_{N}^{(ext)}(\Phi_{A}^{(0)}+X) ,
\eeq
where the first term in the brackets, with $\Phi_{A}^{(0)}$
given by (37), represents the
contribution from the term with
$ j=1 $ (this is the contribution (34), which survives in
the low-energy limit) and the second term represents all
the terms with
$ j>1 $. It is given by an integral
\[
X=A(A-1){\mbox Re}\ \int d^{2}b\int d\xi
e^{-i\Delta\xi}
\int_{-\infty}^{+\infty}dz\rho(b,z)\rho(b,z+\xi)\]\beq
 \left(2\tia+\eta\tis+\tia T(b,z,z+\xi)(A\tia+(A-1)\eta\tis )-
\eta\tis(2\tia+\eta\tis)T (b,z+\xi)\right)(1+w)^{A-3} ,
 \eeq
where we have put
\beq
w=\tia T(b,z,z+\xi)-\eta\tis T(b,z+\xi)
\eeq
and $ \eta=1-\varepsilon $.

A similar result holds for the internal contribution from the target
nucleus:
\beq
I_{A}^{(int,A)}(x)=I_N^{(int,N)}(x)(\Phi_{A}^{(0)}+X) .
\eeq

 For the internal
contribution from the projectile  we obtain
\beq
I_{A}^{(int)}=I_N^{(int,p)}(\Phi_{A}^{(0)}+Y) ,
\eeq
where again the first term comes from
$ j=1 $ and is the one that survives at low energies and the
rest comes from
$ j>1 $ and is given by an integral
\[
Y=-2A(A-1){\mbox Re}\ \tia
\frac{(\tia+\eta\tis)}{\varepsilon\tis}\]\beq \int d^{2}b\int d\xi
e^{-i\Delta\xi}
\int_{-\infty}^{+\infty}dz\rho(b,z)\rho(b,z+\xi)
(1+w)^{A-2} \ \ \ .
\eeq

The calculations although rather cumbersome are now feasible. We have taken a
standard Saxon-Woods form for the nuclear density $\rho(b,z)$ and replaced all
$(1+\tia T)^{A}$ by $\exp (A\tia T)$.

\section{Numerical results and discussion}

In this Section we present our results for $pPb$
collisions at several energies. For onium
production we take $\varepsilon = 0.001$
and for open charm and bottom $\varepsilon =
0.999$.
In the calculations we take the amplitudes $a$ and $\widetilde{a}$ purely
imaginary. First, we restrict ourselves to the nuclear effects due to the
rescattering of the heavy system, i.e. we take only linear terms in $a$ or
$\sigma$.

We start with the asymptotic formulae of Sect. 2. Neglecting the EMC effect
at low $x$
\beq
\Phi_A\equiv AR_A\simeq A\int d^2 b T_A (b)\exp
\left(-\frac{1}{2}\tilde{\sigma}
AT_A (b)\right).
\eeq
 We present our
results in terms of $A_{eff} = I_{pPb}/I_{pN}$.
The three parts of $A_{eff}$,
 the external, internal from the projectile and
internal from the target, are given by
\beq
A_{eff}^{(ext)}=\frac{AR_A}{1+r_1 +r_2 },\
A_{eff}^{(int,p)}=
\frac{\sigma^{(\varepsilon)}_{\Psi A}/
\varepsilon\tilde{\sigma}}{1+1/r_1 +r_2 /r_1},\
A_{eff}^{(int,A)}=\frac{AR_A }{1+1/r_2 +r_1 /r_2},
\eeq
where $r_1$ and $r_2$ are ratios to the external part of the internal ones
contributed by the
 projectile and target, respectively, for a single nucleon as a target.
These ratios are shown in Fig. 3 for open charm ($\varepsilon =0.999$) taking
$\tilde{\sigma}=$7 mb at
$s=60\  GeV^2$
in accordance with the data
\cite{1.,10.}\footnote{This is the center of mass energy of the collision
$\Psi$-$N$ for experiments at  $\sqrt{s}$=20 GeV and $x_F$=0.}.
For the structure functions
$F_{p,N}$ we have  taken the GRV LO parametrization \cite{11.}
and the structure
function $F_{\Psi/p}$ has been calculated using Eq. (10). One observes that
$r_1$ is very small for all $x_F>0$, whereas $r_2$ grows with $x_F$
and becomes greater than unity for $x_F>0.4\div 0.5$.
From (51) we then conclude that
the internal contribution from the projectile proton is
very small for all $x_F$ and can safely be neglected. The internal contribution
from the nuclear target, though also very small in the central region,
becomes dominant at high $x_F$, since the rest vanishes as $x_F\rightarrow 1$.
The total $A_{eff}$ is given by
\beq
A_{eff}\simeq A_{eff}^{(ext)}+A_{eff}^{(int,A)}=AR_A,
\eeq
with the absorptive factor $R_A$ irrespective of the relative weight of the
two remaining contributions.

As commented in the Introduction, this absorptive factor is different from the
probabilistic one. However for $J/\psi$ production this difference is quite
small numerically. One can see this from Table 1, where we present $A_{eff}$
and its three components compared to the probabilistic value $A^{prob}$
at $x_{F}=0$ and 0.5 and various energies. The dependence of $\tilde{\sigma}$
on the energy was taken in accordance with the soft Pomeron picture:
$\tilde{\sigma}\sim s^{0.08}$.

A completely different result follows for open charm production, for which
the probabilistic approach gives no absorption altogether. Our asymptotic
formulas, on the contrary, lead to considerable absorption, as presented in
Table 2. However these results  can only be trusted
at very high energies, as we shall presently see.

To see the finite energy effects we calculated the same quantities
using our formalism of Section 3. The results are presented in Tables 3 and 4
for $J/\psi$ and open charm, respectively. Comparing with Tables 1 and 2,
we observe that finite energy effects are much stronger for the open charm than
for the hidden one. Indeed, for open charm at low energies the finite energy
corrections make the absorption quite small ($\alpha\sim 0.98$ in
the A-dependence $A^\alpha$), bringing the results in accordance with the
experimental data \cite{4.}. This effect only dies out at $\sqrt{s}\geq 200$
GeV, when
our model predicts an absorption for the open charm of the same order as for
the hidden one. For $J/\psi$ the finite energy effects turn out to be rather
small. They also reduce absorption,
but the effect is insignificant ($\sim$ 3\%).

As to the $x_F$-dependence, in all cases absorption grows with $x_F$ due to
the growth with energy of the $\Psi$-$N$ cross-section, although this
growth is rather mild. However it is strengthened if one takes into account
the shadowing corrections to the nuclear structure function, which are quite
large at very small $x_2$ relevant for heavy flavour production at large
$x_F$ and $s$.
This has been done in a simplified manner multiplying our results by
the absorption factor due to nuclear corrections to structure functions taken
from Refs. \cite{12.} and \cite{noso}.
The results are shown in the
last but one columns of Tables 1 and 2 and in the last columns of Tables
3-6. We see that the $x_F$ dependence
of the EMC effect is quite important. Once it is taken into account, the
resulting $x_F$-dependence of $J/\psi$ suppression turns out to be consistent
with the experimental one \cite{5.,noso}. Note, however, that at the two
lower energies of the Tables, the experimentally observed $A$-dependence of
Drell-Yan pair production is very close to $A^1$ at all values of $x_F$. Thus,
a detailed check of this $A$-dependence, with our formalism for the EMC effect,
is needed before we can claim to have a consistent explanation of the
$x_F$-dependence of $J/\psi$ suppression. This interesting point
is,
however, beyond the scope of the present work.

In Tables 5 and 6  we present the same results for hidden and
open bottom production respectively
taking $\widetilde{\sigma} = 2$ mb at $\sqrt{s}=20$ GeV  and assuming the
same energy dependence as in the case of charm production.
The $A$-dependence of $\Upsilon$
at LHC energies and $x_F$=0
corresponds to $\alpha$ = 0.96, which value is lowered to
$\alpha$ = 0.91 when  shadowing in
the nuclear structure functions is introduced.

Finally  results  for $J/\psi$ production at $x_F=0$ in $Pb$-$Pb$ collisions
 are presented in Table 7.  The
difference between
the asymptotic formula and the probabilistic one is larger than
for $pA$ but still small.

\section{Conclusions}

The probabilistic formula used up to now to describe heavy flavour
production off
nuclei has been generalized to all energies using a quantum field theoretical
approach. For $J/\psi$ and $\Upsilon$ production it gives practically the same
results as the probabilistic formula up to $\sqrt{s} \simeq 6$ TeV.
For open heavy
flavour production we predict nuclear absorption for $\sqrt{s} \ \gsim$ 40 GeV.
For $\sqrt{s}\geq$200 GeV this absorption
turns out to be almost the same as the suppression of the $J/\psi$.
Our formalism 
also predicts an increase of the $J/\psi$ suppression with increasing
$x_F$.

\vskip 1.0 cm
{\bf Acknowledgements}

We are grateful to Yu. M. Shabelski and A. B. Kaidalov
for fruitful discussions and to
Direcci\'on
General de Pol\'\i tica Cient\'\i fica and CICYT of Spain for financial
support
under
contract AEN96-1673.
N. A. and C. A. S. also thank Xunta de Galicia for financial support.

\pagebreak

\centerline{{\large\bf Figure captions}}
\vskip 2cm
{\bf Fig. 1a.} Low-energy heavy flavour production amplitude with
rescattering.
\vskip 1cm

{\bf Fig. 1b.}
High-energy heavy flavour production amplitude with rescattering.
\vskip 1cm

{\bf Fig. 2a.} Factorized diagram for heavy flavour production without
rescattering.
\vskip 1cm

{\bf Fig. 2b.} Same as Fig. 2a but with rescattering of the heavy system
with partons from the nucleus $A$.
\vskip 1cm

{\bf Fig. 3.} Ratios of the internal parts over the external one for projectil
(dashed line) and target (solid line) contributions for open charm.

\newpage

\centerline{{\large\bf Tables}}
\vskip 2 cm

\begin{table}[hbt]
\begin{center}
\begin{tabular}{|c||c|c|c|c|c|c|}\hline
  $\sqrt{s}, GeV$ & $A^{ext}_{eff}$ & $A^{int,p}_{eff}$ &
$A^{int,A}_{eff}$ &
$A^{tot}_{eff}$ & $A^{tot+EMC}_{eff}$ & $A^{prob}$ \\
  \hline
\multicolumn{7}{|c|} {$x_F$=0} \\ 
  \hline
  20      & 128.9 & 3.8 & 0.016 & 132.7 & 144.6 & 134.4 \\
  39      & 125.7 & 2.5 & 0.010 & 128.2 & 134.5 & 131.7 \\
  200     & 117.6 & 1.0 & 0.004 & 118.6 & 101.8 & 124.7 \\
  6000    & 99.4 &  0.3 & 0.001 & 99.8 & 71.0 & 109.5 \\
\hline
\multicolumn{7}{|c|} {$x_F$=0.5} \\ 
  \hline
  20      & 122.3 & 0.50 & 0.4 & 123.2 & 117.2 & 129.1 \\
  39      & 115.9 & 0.27 & 0.3 & 116.5 & 96.9 & 123.6 \\
  200     & 98.6 & 0.08 & 0.2 & 98.9 & 70.3 & 109.1 \\
  6000    & 61.4 & 0.01 & 0.2 & 61.6 & 43.1 & 78.6 \\
\hline
\end{tabular}
\caption{Effective atomic numbers for $J/\Psi$ production in
$pPb$ collisions with asymptotic
formulae.}
\end{center}
\end{table}
\vskip 2cm
\begin{table}[hbt]
\begin{center}
\begin{tabular}{|c||c|c|c|c|c|c|}\hline
  $\sqrt{s}, GeV$ & $A^{ext}_{eff}$ & $A^{int,p}_{eff}$ &
$A^{int,A}_{eff}$ &
$A^{tot}_{eff}$ & $A^{tot+EMC}_{eff}$ & $A^{prob}$ \\
  \hline
\multicolumn{7}{|c|} {$x_F$=0} \\ 
  \hline
  20      & 103.1 & 16.5 & 12.9 & 132.5 & 144.4 & 207.9\\
  39      & 108.3 & 11.3 & 8.7 & 128.3 & 131.9 & 207.9 \\
  200     & 110.6 & 4.7 & 3.5 & 118.8 & 101.9 & 207.9 \\
  6000    & 97.9 & 1.2 & 0.8 & 99.9 & 71.0 & 207.8 \\
\hline
\multicolumn{7}{|c|} {$x_F$=0.5} \\ 
  \hline
  20      & 28.3 & 0.58 & 93.9 & 122.8 & 116.8 & 207.9 \\
  39      & 32.5 & 0.36 & 83.4 & 116.3 & 96.8 & 207.9 \\
  200     & 28.5 & 0.09 & 70.4 & 98.9 & 70.3 & 207.8 \\
  6000    & 17.1 & 0.01 & 44.4 & 61.6 & 43.1 & 207.7 \\
\hline
\end{tabular}
\caption{Effective atomic numbers for open charm production in
$pPb$ collisions with asymptotic
formulae.}
\end{center}
\end{table}
\vskip 2cm

\begin{table}[hbt]
\begin{center}
\begin{tabular}{|c||c|c|c|c|c|}\hline
  $\sqrt{s}, GeV$ & $A^{ext}_{eff}$ & $A^{int,p}_{eff}$ &
$A^{int,A}_{eff}$ &
$A^{tot}_{eff}$ & $A^{tot+EMC}_{eff}$ \\
  \hline
\multicolumn{6}{|c|} {$x_F$=0} \\ 
  \hline
  20      & 134.9 & 0.1 & 0.017 & 135.0 & 147.2\\
  39      & 133.2 & 0.6 & 0.011 & 133.8 & 137.5\\
  200     & 118.4 & 1.0 & 0.004 & 119.4 & 102.4\\
  6000    & 99.6 & 0.3 & 0.001 & 99.9 & 71.0\\
\hline
\multicolumn{6}{|c|} {$x_F$=0.5} \\ 
  \hline
  20      & 126.4 & 0.33 & 0.42 & 127.2 & 121.0\\
  39      & 116.5 & 0.26 & 0.30 & 117.0 & 97.3\\
  200     & 98.7 & 0.08 & 0.24 & 99.1 & 70.5\\
  6000    & 61.4 & 0.01 & 0.16 & 61.6 & 43.1\\
\hline
\end{tabular}
\caption{Finite Energy effective atomic numbers for $J/\Psi$ production in
$pPb$ collisions.}
\end{center}
\end{table}
\vskip 2cm

\begin{table}[hbt]
\begin{center}
\begin{tabular}{|c||c|c|c|c|c|}\hline
  $\sqrt{s}, GeV$ & $A^{ext}_{eff}$ & $A^{int,p}_{eff}$ &
$A^{int,A}_{eff}$ &
$A^{tot}_{eff}$ & $A^{tot+EMC}_{eff}$ \\
  \hline
\multicolumn{6}{|c|} {$x_F$=0} \\ 
  \hline
  20      & 162.7 & 20.6 & 20.4 & 203.7 & 222.0 \\
  39      & 155.8 & 13.5 & 12.5 & 181.8 & 186.9\\
  200     & 114.1 & 4.7 & 3.6 & 122.5 & 105.1\\
  6000    & 97.8 & 1.2 & 0.8 & 99.8 & 71.0\\
\hline
\multicolumn{6}{|c|} {$x_F$=0.5} \\ 
  \hline
  20      & 33.8 & 0.63 & 112.2 & 146.7 & 139.5 \\
  39      & 33.2 & 0.36 & 85.1 & 118.6 & 98.7\\
  200     & 28.5 & 0.09 & 70.3 & 98.9 & 70.3\\
  6000    & 17.0 & 0.01 & 44.1 & 61.1 & 42.7\\
\hline
\end{tabular}
\caption{Finite Energy effective atomic numbers for open charm production in
$pPb$ collisions.}
\end{center}
\end{table}
\vskip 2cm

\begin{table}[hbt]
\begin{center}
\begin{tabular}{|c||c|c|c|c|c|}\hline
  $\sqrt{s}, GeV$ & $A^{ext}_{eff}$ & $A^{int,p}_{eff}$ &
$A^{int,A}_{eff}$ &
$A^{tot}_{eff}$ & $A^{tot+EMC}_{eff}$ \\
  \hline
\multicolumn{6}{|c|} {$x_F$=0} \\ 
  \hline

      20 & 181.5 & 0.02 & 0.006 & 181.5 & 197.1\\
      39 & 180.3 & 0.06 & 0.004 & 180.4 & 186.5\\
     200 & 175.9 & 0.12 & 0.002 & 176.0 & 157.7\\
    6000 & 166.9 & 0.04 & 0.000 & 166.9 & 127.5\\

\hline
\multicolumn{6}{|c|} {$x_F$=0.5} \\ 
  \hline

      20  &   178.4  &     0.040  &     0.2 &    178.7 & 173.8\\ 
      39  &   175.1  &     0.034  &     0.1 &    175.3 & 153.6 \\
     200  &   166.5  &     0.012  &     0.1 &    166.6 & 126.7\\
    6000  &   142.5  &     0.003  &     0.1 &    142.6 & 82.8\\
\hline
\end{tabular}
\caption{Effective atomic numbers for $\Upsilon$ production in
$pPb$ collisions.} 
\end{center}
\end{table}
\vskip 2cm

\begin{table}[hbt]
\begin{center}
\begin{tabular}{|c||c|c|c|c|c|}\hline
  $\sqrt{s}, GeV$ & $A^{ext}_{eff}$ & $A^{int,p}_{eff}$ &
$A^{int,A}_{eff}$ &
$A^{tot}_{eff}$ & $A^{tot+EMC}_{eff}$ \\
  \hline
\multicolumn{6}{|c|} {$x_F$=0} \\ 
  \hline

      20 &    193.3  &     7.0  &     6.9  &   207.2 & 224.4 \\
      39 &    192.5  &     4.5  &     4.4  &   201.4 & 208.2\\
     200 &    174.4  &     1.7  &     1.5  &   177.6 & 159.4\\
    6000 &    166.2  &     0.4  &     0.3  &   167.1 & 127.6\\

\hline
\multicolumn{6}{|c|} {$x_F$=0.5} \\ 
  \hline

      20 &     96.5  &     0.45  &    91.4  &   188.3 & 186.2\\ 
      39 &    101.6  &     0.26  &    74.4  &   176.3 & 156.6\\
     200 &     97.7  &     0.07  &    69.0  &   166.7 & 127.6\\
    6000 &     81.9  &     0.01  &    60.7  &   142.7 & 83.4\\
\hline
\end{tabular}
\caption{Effective atomic numbers for open bottom production in 
$pPb$ collisions.}
\end{center}
\end{table}
\vskip 2cm

\begin{table}[hbt]
\begin{center}
\begin{tabular}{|c||c|c|c|c|c|}\hline
\multicolumn{6}{|c|} {Asymptotic Energies} \\ 
  \hline
  $\sqrt{s}, GeV$ & $A^{ext}_{eff}$ & $A^{int,p}_{eff}$ &
$A^{int,A}_{eff}$ &
$A^{tot}_{eff}$ & $A^{prob}$ \\
  \hline

      20 & 16625.5  & 492.8 & 2.1 & 17120.4 & 18078.1  \\
      39 & 15809.0  & 316.6 & 1.3 & 16126.9 & 17341.2 \\
     200 & 13825.1  & 123.0 & 0.5 & 13948.6 & 15552.8 \\
    6000 & 9896.7  & 29.8 & 0.1 & 9926.6 & 11997.9 \\

\hline
\end{tabular}

\vskip 1cm
\begin{tabular}{|c||c|c|c|c|}\hline
\multicolumn{5}{|c|} {Finite Energies} \\ 
  \hline
  $\sqrt{s}, GeV$ & $A^{ext}_{eff}$ & $A^{int,p}_{eff}$ &
$A^{int,A}_{eff}$ &
$A^{tot}_{eff}$ \\
  \hline

      20 & 18316.1  & 13.2  & 2.3 & 18231.6 \\ 
      39 & 17754.0  & 74.2  & 1.5 & 17829.7 \\
     200 & 14014.1  & 117.3  & 0.5  & 14131.9 \\
    6000 & 9913.2  & 29.9  & 0.1  & 9943.2 \\
\hline
\end{tabular}
\caption{Effective atomic numbers for $J/\Psi$ production in 
$PbPb$ collisions at $x_F$=0.}
\end{center}
\end{table}
 

\newpage

\begin{figure}[hbt]
\begin{center}
\mbox{\epsfig{file=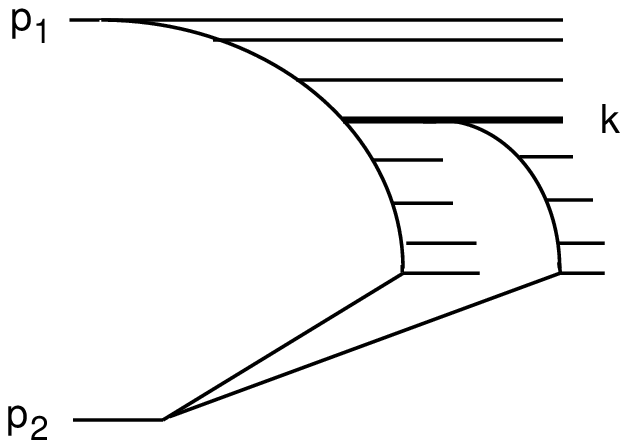,height=8.0cm}}
\large{\bf Fig. 1a.}
\end{center}
\end{figure}

\begin{figure}[hbt]
\begin{center}
\mbox{\epsfig{file=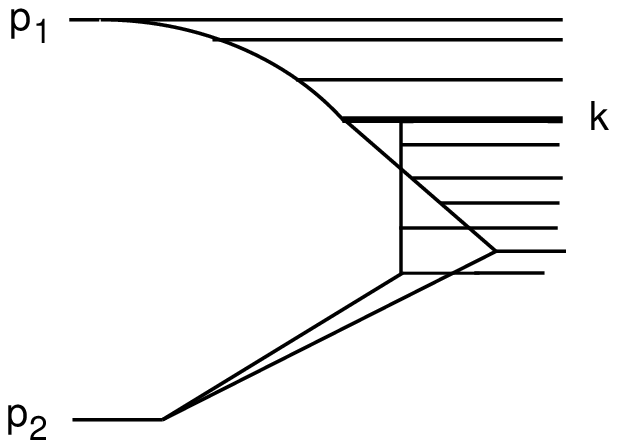,height=8.0cm}}
\large{\bf Fig. 1b.}
\end{center}
\end{figure}

\newpage
\begin{figure}[hbt]
\begin{center}
\mbox{\epsfig{file=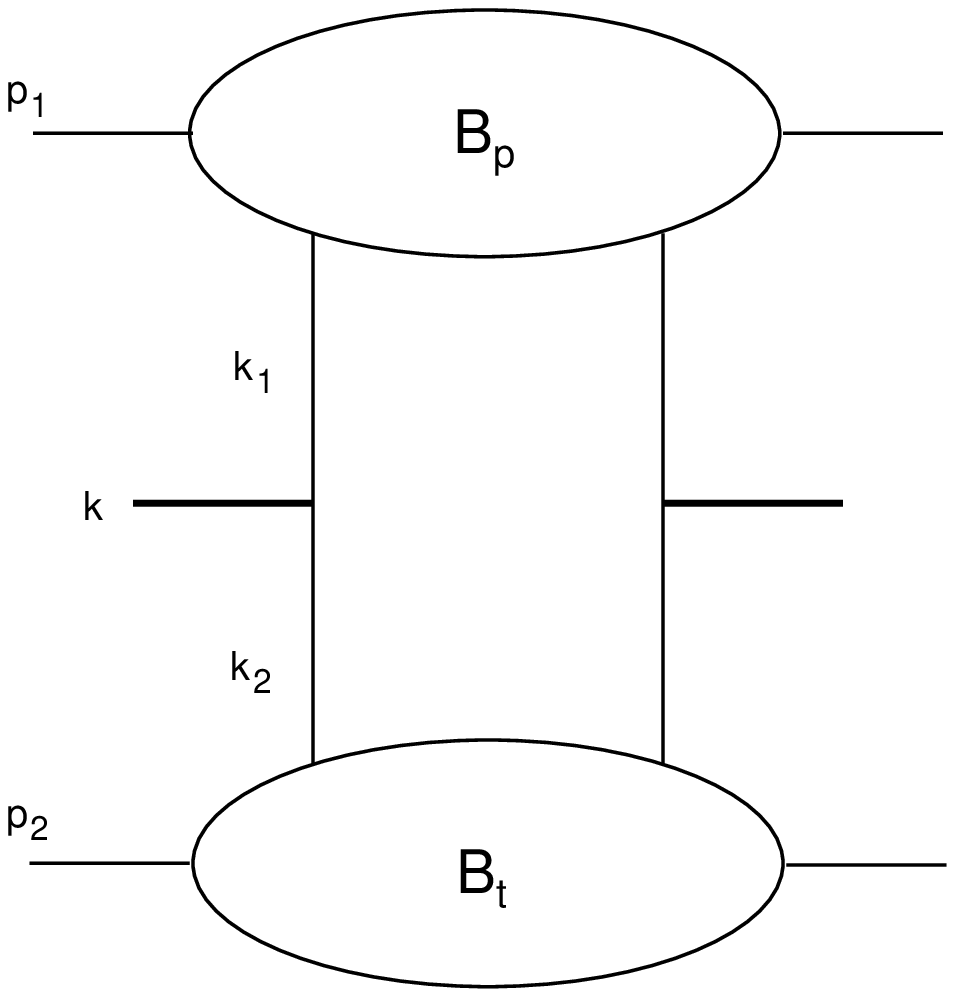,height=8.0cm}}
\large{\bf Fig. 2a.}
\end{center}
\end{figure}

\begin{figure}[hbt]
\begin{center}
\mbox{\epsfig{file=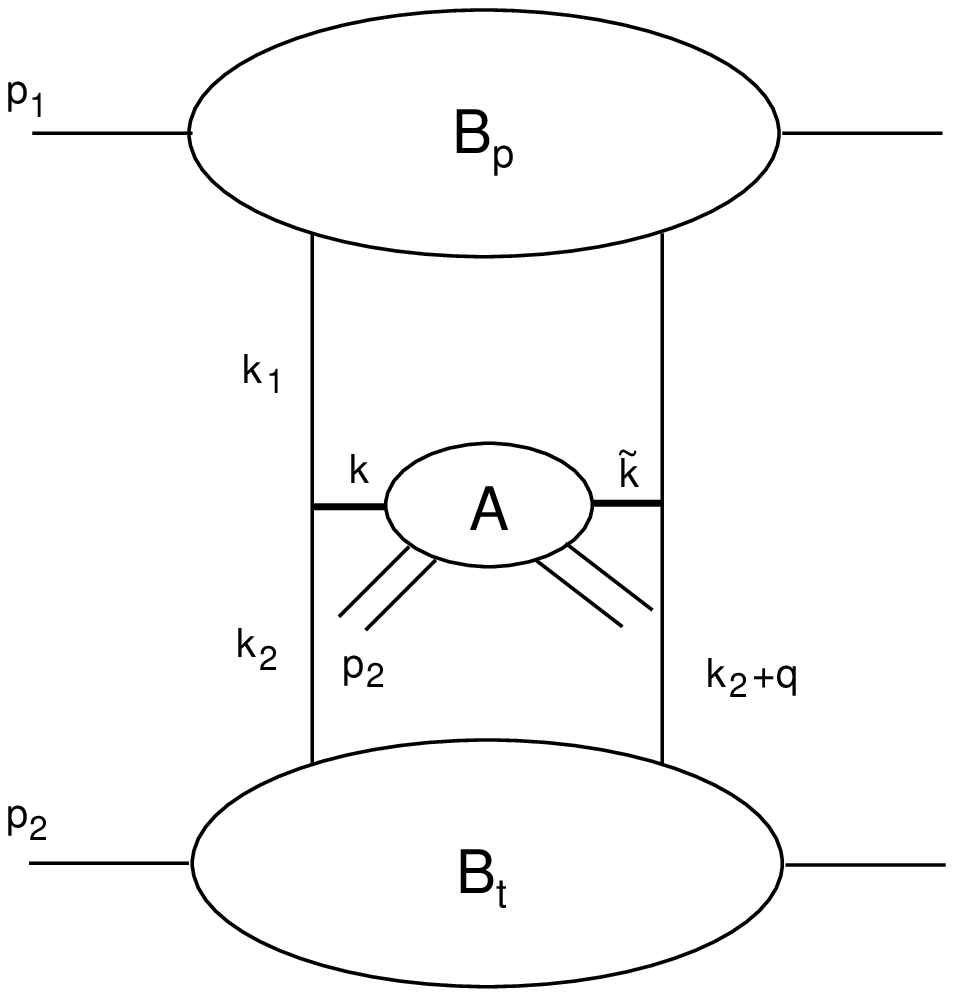,height=8.0cm}}
\large{\bf Fig. 2b.}
\end{center}
\end{figure}

\newpage
\begin{figure}[hbt]
\begin{center}
\mbox{\epsfig{file=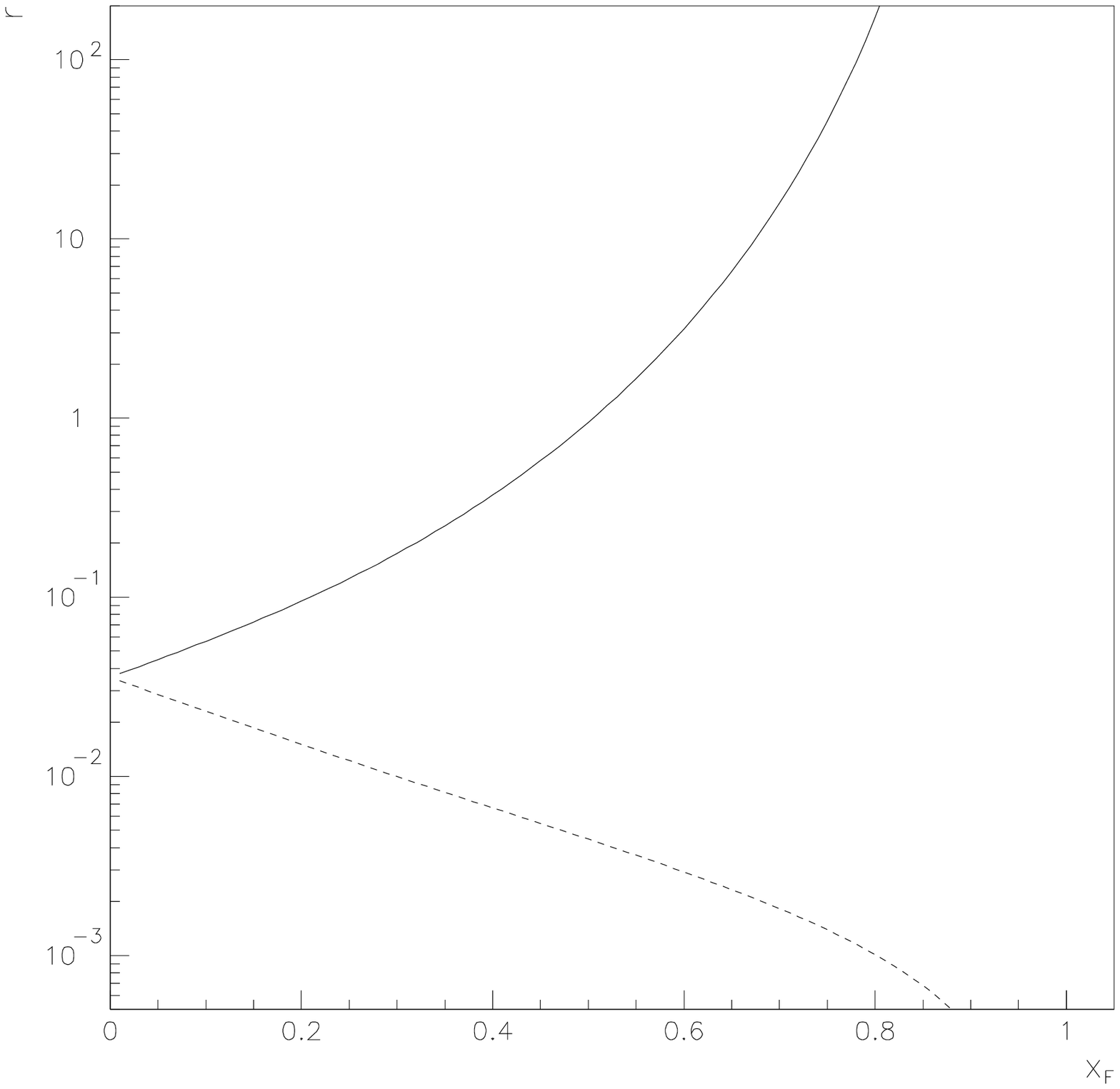,height=14cm}}
\end{center}
\begin{center}
\large{\bf Fig. 3.}
\end{center}
\end{figure}

\end{document}